\documentstyle[psfig]{mn}

\newif\ifAMStwofonts

\ifoldfss
  \ifCUPmtlplainloaded \else
    \NewTextAlphabet{textbfit} {cmbxti10} {}
    \NewTextAlphabet{textbfss} {cmssbx10} {}
    \NewMathAlphabet{mathbfit} {cmbxti10} {} 
    \NewMathAlphabet{mathbfss} {cmssbx10} {} 
  \fi
  \ifAMStwofonts
    \ifCUPmtlplainloaded \else
      \NewSymbolFont{upmath} {eurm10}
      \NewSymbolFont{AMSa} {msam10}
      \NewMathSymbol{\upi}     {0}{upmath}{19}
      \NewMathSymbol{\umu}     {0}{upmath}{16}
      \NewMathSymbol{\upartial}{0}{upmath}{40}
      \NewMathSymbol{\leqslant}{3}{AMSa}{36}
      \NewMathSymbol{\geqslant}{3}{AMSa}{3E}

    \fi
  \fi
\fi 

\ifnfssone
  \newmathalphabet{\mathit}
  \addtoversion{normal}{\mathit}{cmr}{m}{it}
  \addtoversion{bold}{\mathit}{cmr}{bx}{it}
  \newmathalphabet{\mathbfit} 
  \addtoversion{normal}{\mathbfit}{cmr}{bx}{it}
  \addtoversion{bold}{\mathbfit}{cmr}{bx}{it}
  \newmathalphabet{\mathbfss} 
  \addtoversion{normal}{\mathbfss}{cmss}{bx}{n}
  \addtoversion{bold}{\mathbfss}{cmss}{bx}{n}
  \ifAMStwofonts
    \ifCUPmtlplainloaded \else
      \UseAMStwoboldmath
      \makeatletter
      \new@mathgroup\upmath@group
      \define@mathgroup\mv@normal\upmath@group{eur}{m}{n}
      \define@mathgroup\mv@bold\upmath@group{eur}{b}{n}
      \edef\UPM{\hexnumber\upmath@group}
      \new@mathgroup\amsa@group
      \define@mathgroup\mv@normal\amsa@group{msa}{m}{n}
      \define@mathgroup\mv@bold\amsa@group{msa}{m}{n}
      \edef\AMSa{\hexnumber\amsa@group}
      \makeatother
      \mathchardef\upi="0\UPM19
      \mathchardef\umu="0\UPM16
      \mathchardef\upartial="0\UPM40
      \mathchardef\leqslant="3\AMSa36
      \mathchardef\geqslant="3\AMSa3E
    \fi
  \fi
\fi 

\ifnfsstwo
  \DeclareMathAlphabet{\mathbfit}{OT1}{cmr}{bx}{it}
  \SetMathAlphabet\mathbfit{bold}{OT1}{cmr}{bx}{it}
  \DeclareMathAlphabet{\mathbfss}{OT1}{cmss}{bx}{n}
  \SetMathAlphabet\mathbfss{bold}{OT1}{cmss}{bx}{n}
  \ifAMStwofonts
    \ifCUPmtlplainloaded \else
      \DeclareSymbolFont{UPM}{U}{eur}{m}{n}
      \SetSymbolFont{UPM}{bold}{U}{eur}{b}{n}
      \DeclareSymbolFont{AMSa}{U}{msa}{m}{n}
      \DeclareMathSymbol{\upi}{0}{UPM}{"19}
      \DeclareMathSymbol{\umu}{0}{UPM}{"16}
      \DeclareMathSymbol{\upartial}{0}{UPM}{"40}
      \DeclareMathSymbol{\leqslant}{3}{AMSa}{"36}
      \DeclareMathSymbol{\geqslant}{3}{AMSa}{"3E}
    \fi
  \fi
\fi 

\ifCUPmtlplainloaded \else
  \ifAMStwofonts \else 
    \def\upi{\pi}
    \def\umu{\mu}
    \def\upartial{\partial}
  \fi
\fi

\title{Long-term Properties of Accretion Disks in X-ray Binaries: I. the variable third period in SMC X-1}
\author[Clarkson, Charles, Coe, Laycock, Tout, Wilson]
       {W.I.~Clarkson$^{1}$,P. A.~Charles$^{1}$, M. J.~Coe$^{1}$, S.~Laycock$^{1}$, M.D.~Tout$^{1}$, C.A.~Wilson$^{2}$ \\ 
       1. Department of Physics and Astronomy, Southampton University, SO17 1BJ, UK \\
       2. NASA/Marshall Space Flight Center, Huntsville, AL 35805, USA}

\date{Accepted .
      Received ;
      in original form 
      }

\pagerange{\pageref{firstpage}--\pageref{lastpage}}
\pubyear{2001}

\begin{document}

\maketitle

\label{firstpage}

\begin{abstract}

Long term X-ray monitoring data from the RXTE ASM and CGRO BATSE
reveal that the third (superorbital) period in SMC X-1 is not
constant, but varies between 40-60 days. A dynamic power spectrum
analysis indicates that the third period has been present continuously
throughout the five years of ASM observations. This period changed
smoothly from 60 days to 45 days and then returned to its former
value, on a timescale of approximately 1600 days. During the nearly 4
years of overlap between the CGRO \& RXTE missions, the simultaneous
BATSE hard X-ray data confirm and extend this variation in SMC
X-1. Our discovery of such an instability in the superorbital period
of SMC X-1 is interpreted in the context of recent theoretical studies
of warped, precessing accretion disks.  We find that the behaviour of
SMC X-1 is consistent with a radiation-driven warping model.

\end{abstract}

\begin{keywords}
stars - X-rays: binaries :pulsars :accretion disks :precession periods
\end{keywords}

\section{Introduction}

The X-ray lightcurves of persistent neutron star X-ray binaries
exhibit periodicities which give information about physical mechanisms
at work in the systems. Systems with high magnetic fields can show
X-ray pulsations on timescales from milliseconds to seconds, due to
channeling of accreting matter onto small emitting regions on the
neutron star \cite{wnp}. Also found are quasi - periodic oscillations,
on disk - magnetosphere interaction timescales ranging from
milliseconds to fractions of a second \cite{vdk}. Orbital modulations
are also seen on timescales ranging from a few tens of minutes to a
few tens of days. The orbit manifests itself in the X-ray as either a
change in mass transfer (and therefore emission) in an eccentric
orbit, or a change in the character or magnitude of absorption, in
some cases eclipses by the donor \cite{vpm}.

\subsection{Superorbital Periods}

However, in a small number of sources with high or low-mass companions
(e.g. SMC X-1, Her X-1, LMC X-4) a third (or ``superorbital'') period
is also present, typically on a scale of tens of days.  Early in the
history of X-ray astronomy an explanation for the behaviour of the
LMXB Her X-1 was suggested (Petterson 1977) which involved a disk that
was tilted and/or warped by the intense radiation pressure from the
central X-ray source, thereby causing it to precess with the observed
long period. In doing so it can periodically obscure the central
source, giving rise to a modulation of the X-ray flux at the
precession period.  This model has been the subject of renewed
theoretical examination in recent years (see e.g. Wijers and Pringle
1999). Other models for superorbital periods have been proposed,
including precession of the magnetic axis of the neutron star
\cite{tru} and a hierarchical third body \cite{chg}.

It should also be noted that superorbital periods on timescales as
long as hundreds of days have been reported in a number of X-ray
sources (see e.g. Smale and Lochner 1992), namely those bright enough
to be detectable by the first generation of all-sky monitors (such as
those on Vela-5B and Ariel 5). However, these modulations are
distinctly ``quasi-periodic'' in that even long-term monitoring
datasets do not yield precise periods in their power spectra, but a
broad peak, often superposed on ``red-noise'' power.  Examples of
these include Cyg X-2 ($P\sim$70-80d; Paul, Kitamoto \& Makino 2000),
GX354-0 ($P\sim$70d; Kong, Charles \& Kuulkers 1998, X1916-051
($P\sim$83d; Homer et al. 2001) and X1820-30 ($P\sim$200d; Chou \&
Grindlay 2001), although in the latter case the long period is
sufficiently stable that it has been suggested
\cite{chg} that X1820-30 could be a triple system.  Most of these
modulations are rather different from those exhibited by SMC X-1, Her
X-1 and LMC X-4 in that they are much lower amplitude and do not
appear to be due to varying obscuration. It has been customary to also
interpret these superorbital periods as due to a precessing, possibly
warped accretion disc, with the lower amplitudes due to lower
inclinations.

Clearly the detailed investigation of these long-period phenomena is
an observational challenge in that it really requires regular,
systematic monitoring of the X-ray flux, preferably with a single
instrument in each energy range. In the energy range 1.3 - 12.1 keV,
the RXTE All-Sky Monitor (ASM) provides a continuous dataset of all
bright X-ray sources, that has been continuously compiled over a
period of more than six years. Up to energies of 1 MeV, the Burst And
Transient Source Experiment (BATSE) has provided a continuous dataset
over a period of almost a decade. The advent of such long-term
continuous datasets has made the work we describe here possible. This
paper is the first in a series that examines the nature and properties
of these superorbital variations.

\subsection{SMC X-1}

SMC X-1 is a massive X-ray binary system consisting of a 1.6 $\pm$ 0.1
M$_\odot$ neutron star (an X-ray pulsar) and a high mass (17.2 $\pm$
0.6 M$_\odot$) B0 I optical companion, known as Sk 160 (Reynolds et al
1993).  The spin period of the pulsar is 0.71 s and the orbital period
of the system is 3.89 days, and its orbital period is decaying on a
timescale of $10^5$ years (see Reynolds et al. (1993) and references
therein). SMC X-1 also has a superorbital period of approximately 60
days believed to be the consequence of a precessing, warped, accretion
disk (see Wojdowski at al. (1998) and references therein). This disk
is presumed to periodically occult the X-rays from the central source,
thereby creating a modulation in the X-ray lightcurve in a manner
analogous to that of Her X-1.

A comprehensive X-ray investigation of the SMC X-1 system has been
presented by Wojdowski et al \shortcite{woj} which examines all of the
available data up to JD 24450660 (early 1998). As well as showing a
superorbital period of $\sim$ 60 days, they demonstrate that the X-ray
source has been continously active for at least the last 30 years and
that the mass transfer rate to the neutron star has been roughly
constant over this time. The latter is inferred from the fact that the
pulsar has been steadily spinning-up on a timescale of hundreds of
years. Based on the placement of pulse and eclipse profiles within the
superorbital cycle, it was suggested that the luminosity and spectrum
during the (orbital) eclipses are unaffected by the long cycle. It was
thus argued that no intrinsic source variation could be responsible
for the 60 day modulation, since the superorbital variation was
present only in the out-of-eclipse data. Periodic occultation of the
pulsar by the rim of a warped accretion disk, precessing under the
gravity of the giant companion was shown to be consistent with the
observations. 

However, whilst confirming the presence of the superorbital
modulation, Wojdowski et al also noted that it was not steady, but
appeared to vary between 50 and 60 days. The identification of
pointings with phase in the superorbital cycle could therefore not be
made independently of the pointed lightcurves themselves. Instead,
timing in the superorbital cycle was performed by creating a
consistent scenario, then fitting the pointings to that scenario. We
therefore believe that the elimination of intrinsic luminosity
variation is open to question.

\begin{figure}
\begin{center}
\psfig{file=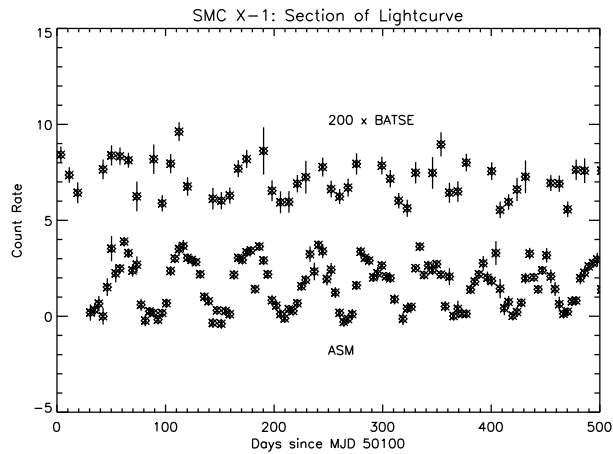,width=8cm}
\caption{Section of the ASM and BATSE datasets for SMC X-1. Data has been binned to 7.78 days, or two orbital cycles, per bin.}
\label{fig:lcurve_section}
\end{center}
\end{figure}

In this work we present the results of an analysis of both ASM and
BATSE data that allows the variation in the third period to be
followed continuously over more than a decade. We also extend the
timeline of observations by 3 years since Wojdowski et al
\shortcite{woj}, which allows us to see that the variation of the
third period might itself be periodic, as suggested by Ribo et al
(2001). We show that this variation in the long period must be due to
interaction of modes in a warped, precessing accretion disk.

\section[]{RXTE Observations}

Launched in 1996, the Rossi X-Ray Timing Explorer (RXTE) carries an
All Sky Monitor (ASM), which gives continuous coverage of the entire
sky.  Typically 5-10 readings - called ``dwells'' - are taken of each
of a list of sources per day, lasting about 90 seconds per
dwell. Timing information is provided to within a thousandth of a day,
as well as crude spectral information.  The ASM is sensitive to photon
energies between 1.3 and 12.1 keV, broken into three energy channels
(1.3-3.0~keV, 3.0-5.0~keV and 5.0-12.1~keV). A total of 6.1 years of
data (MJD 50083-52312) from the ASM were used in our analysis.  Since
the X-ray flux of SMC X-1 varies by a factor of $>$10 over the 40-60
day high-low state cycle, the data quality obtained by the ASM is
variable. For all analyses reported in this paper, the ASM data were
selected by background and quality of coverage; all points with
background level above 10 $cs^{-1}$ were rejected. Crude X-ray
spectral information can be obtained from the ASM data by constructing
hardness ratios of the three channels. In this case we define the
ratio to be the sum of the count rate in the two high energy channels
(3.0 - 12.1 keV) to that at low energy (1.3 - 3.0 keV). A section of
the ASM lightcurve can be found in figure \ref{fig:lcurve_section}

\section{BATSE Observations}

The Burst And Transient Source Experiment (BATSE) carried on the
Compton Gamma Ray Observatory (CGRO) provided near continuous
monitoring of SMC X-1 for a period of just over 9 years from JD
24448361 - 24451690. Data were reduced by the Earth Occultation
Technique (EOT) described by Harmon et al \shortcite{har1} and Zhang
et al \shortcite{zha}. By exploiting the Earth's limb as an occulting
mask, individual X-ray sources can be located with an accuracy of
1\degr. During each 90 minute orbit of CGRO, two measurements of the
total 20-100 keV flux from SMC X-1 were made as the source moved into
and out of occultation due to the satellite's motion. Sources at
$|dec| > 41\degr$, such as SMC X-1, undergo intervals, typically
lasting a few days, during each spacecraft precession period where
they are not occulted by the Earth. These intervals manifest
themselves as gaps in the data. To improve signal-to-noise and to
remove effects of spacecraft repointings, daily average count spectra
were generated for each detector by averaging the individual step
measurements for each detector and each energy channel. These count
spectra were then fitted simultaneously for each detector viewing the
source with a photon power law model with a fixed photon index of 3
and a normalization equal to the integrated 20-100 keV flux in photons
$cm^{-2} s^{-1}$ that was forward-folded through the detector response
matrices (Briggs 1995; Pendleton et al. 1995). A fixed photon index
was used because it was not well determined from fits to 1-day average
spectra. The 1-day average 20-100 keV fluxes were used for this
analysis. A section of the BATSE lightcurve can be found in figure
\ref{fig:lcurve_section}.

\section{Dynamic Power Spectrum Analysis}

The ASM and BATSE data were analysed using a `dynamic power spectrum'
approach, designed to allow any changes in the X-ray modulation to be
closely followed.  The Lomb-Scargle periodogram code (Scargle 1982,
1989) implemented in Starlink software PERIOD was adapted in
conjunction with a sliding `data window' to produce power density
spectra (PDS) for a series of overlapping stretches of the time
series. Adjustable parameters in the analysis were the length of the
data window and the amount of time by which the window was shifted
each time to obtain the overlapping stretches of data. The code
accounted for variations in the number of datapoints per interval such
that the power spectrum resolution was identical for each
interval. Choice of data window length was influenced by the expected
minimum number of cycles needed for a significant detection of
modulation due to SMC X-1. Since the `third period' is already
established to be approximately 40-60 days, the data window was set at
4 cycles (200 days).  Of course, a longer window has the effect of
reducing sensitivity to changes on timescales shorter than the window
length.

The amount of shift between consecutive intervals influences the
degree to which changes can be followed, for example setting the shift
to equal the window length, results in zero overlap and no ability to
determine whether an observed change in the PDS was gradual or
sudden. Conversely, excessive overlap reduces the statistical
independence of consecutive intervals. In any case there is no gain in
sensitivity once the barrier of the periodogram resolution is
approached, this being a significant factor when there are only a few
signal cycles in the window. A shift of 50 days was used as this
corresponds to approximately one cycle of the 40-60 day modulation.

The result of the above procedure was an array of periodograms at
identical resolution covering each dataset in many overlapping
intervals. These results were then visualized by constructing
3-dimensional maps in (Time, Frequency, Power) space with spectral
power being represented jointly by grayscale intensity and
contours. The grayscale represented the raw periodograms and the
contours provided a smoothed interpolation over the data.

\begin{figure*}
\begin{center}
\psfig{file=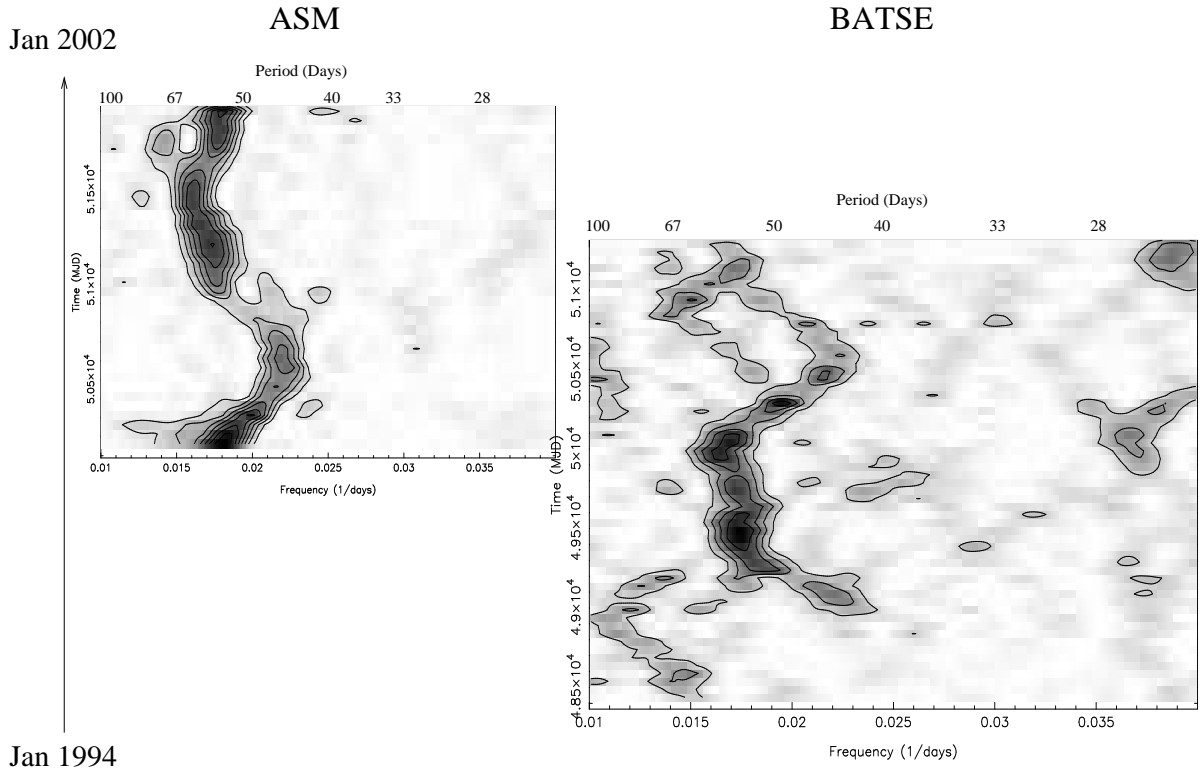,width=17cm,angle=-90}
\caption{
Dynamic power spectrum for the 20-100 keV lightcurve of SMC X-1 observed by BATSE over 9 years (MJD 48361-51690; right) and for the 1.3-12.1 keV lighcturve provided by RXTE over 6 years (MJD 50083-52312; left). Contours spaced at 4-unit intervals of LS power for BATSE and 40-unit intervals for RXTE.
} 
\label{fig:comps}
\end{center}
\end{figure*}

The Dynamic Power Spectra of SMC X-1 in the BATSE and ASM wavebands is
seen in figure \ref{fig:comps}. Not only is there a clear superorbital
period on a period similar to that reported by Wojdowski et al
\shortcite{woj}, but it varies with time, on a scale of $\sim$ 7
years. Simulations show that the smooth shape of variation cannot be
reproduced with a simple replacement of one periodicity at $\sim$ 55
days by another at $\sim$ 47 days, and we believe we are dealing with
variation of a single periodicity. Some additional structure is seen
in the BATSE lightcurve at roughly twice the frequency of the main
structure: this is thought to be a harmonic of the main variation. The
agreement between the datasets suggests that whatever process is
responsible for the variation affects both wavebands.

\subsection{Investigation of systematic errors}

\begin{figure}
\begin{center}
\psfig{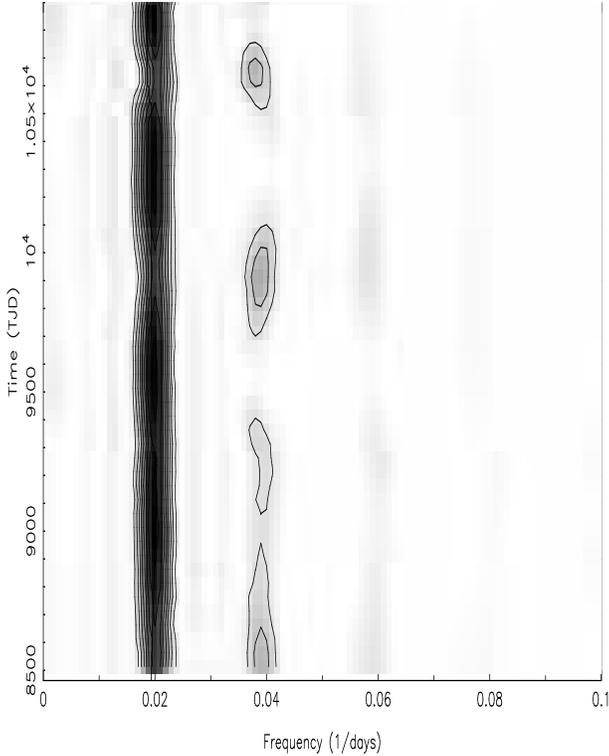}
\caption{Dynamic power spectrum analysis applied to a simulated dataset. 
 The flux values in the BATSE timeseries have been replaced by a 50 day period 
 sinewave combined with gaussian white-noise, to investigate any possible 
  artifacts arising from the timing structure.}
\label{fig:fake}
\end{center}
\end{figure}

To investigate the sensitivity of the analysis to noise and systematic
errors, the analysis was run on artificial lightcurves with the
sampling of the SMC X-1 datasets but with noise superposed. Two cases
were tested for: (1) varying amplitude variations of stable period
with gaussian white noise, (2) white noise only. Red noise, in which
the noise amplitude depends on frequency, was not tested for on the
grounds that there is no evidence for it in the power spectrum of SMC
X-1. Some spectral leakage was found for noisy datasets of known
periodicity (see figure \ref{fig:fake} for an example), identifiable
in that their shape mirrors that of the strongest variation. Pure
white noise produced no periodicities at any time above spectral power
corresponding to roughly 80 \% significance.

In the case of BATSE occultation data, interaction between the
precession period of $\sim$ 51 days and any detected periodicity might
be expected. This possibility was carefully investigated, and found to
produce a negligible effect. The orbital precession period was plotted
for the duration of the BATSE dataset and found not to correlate with
the variation in superorbital period. Furthermore, no bright X-ray
sources are within the Earth Occultation errorbox of 2\degr from SMC
X-1. 

However the key point to note is the close correlation between the ASM
and CGRO results during the period of overlap. This is convincing in
itself because RXTE has a different orbital precession period and the
ASM works on an entirely different principle to BATSE. Data from the
ASM and BATSE provide a 4.4 year stretch of simultaneous coverage in
two separate regions of the X-ray spectrum. During this period of
overlap (MJD 50083-51690), the behaviour of the superorbital period is
closely mirrored in the two, completely independent, datasets. This
can be seen clearly in figure \ref{fig:comps}. Nor is this effect seen
in any other bright X-ray source observed by the CGRO.

\begin{figure}
\begin{center}
\psfig{file=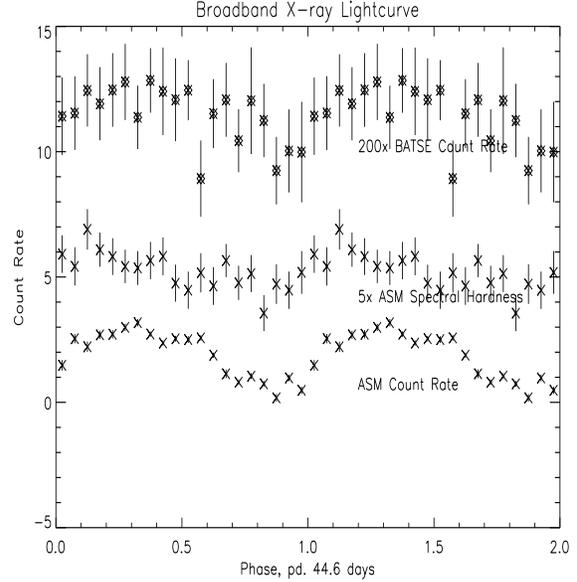,width=8cm,height=8cm}
\caption{ASM countrate, ASM hardness ratio, and scaled BATSE count rate, binned by phase in the quasi-stable superorbital period of 46 days that persists from MJD 50550 - 50800. Spectral Hardness is defined in this paper as the ratio of counts at 3.0 - 12.1 keV to those at 1.3 - 3.0 keV.}
\label{fig:fold_hrat}
\end{center}
\end{figure}

\section{Spectral Behaviour of the Third Period}

We use the spectral information on the variability which is provided
by the multi-channel nature of the ASM dataset, in order to
discriminate between possible superorbital period mechanisms. To
increase signal to noise, binning the data by phase is
desired. However, the variation in superorbital period rules out
folding over the entire dataset. Examination of figure \ref{fig:comps}
shows that over the region MJD 50550 - 50800, the superorbital period
is roughly stable for both ASM and BATSE datasets.

\begin{figure}
\begin{center}
\psfig{file=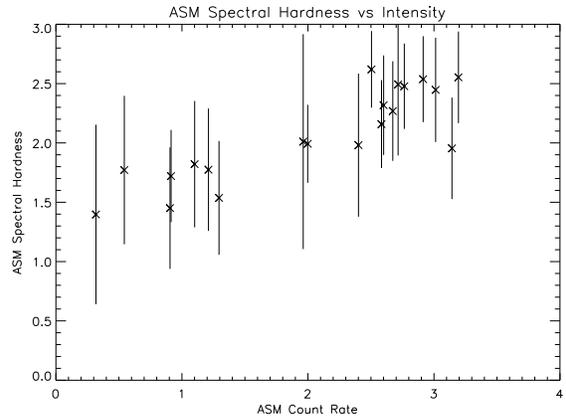,width=8cm}
\caption{ASM spectral hardness plotted against total count rate, for the dataset of figure~\ref{fig:fold_hrat}.}                                 
\label{fig:hrat_counts}
\end{center}
\end{figure}

Figure~\ref{fig:fold_hrat} shows the phase-binned lightcurve of the
SMC X-1 datasets during this interval. We see immediately that the ASM
variations mirror those in of BATSE. Furthermore, the shape of the ASM
total count rate variation is almost exactly reproduced in the
variation of the hardness ratio. The relation between total count rate
and spectral hardness is shown explicitly in
figure~\ref{fig:hrat_counts}.

\begin{figure}
\begin{center}
\psfig{file=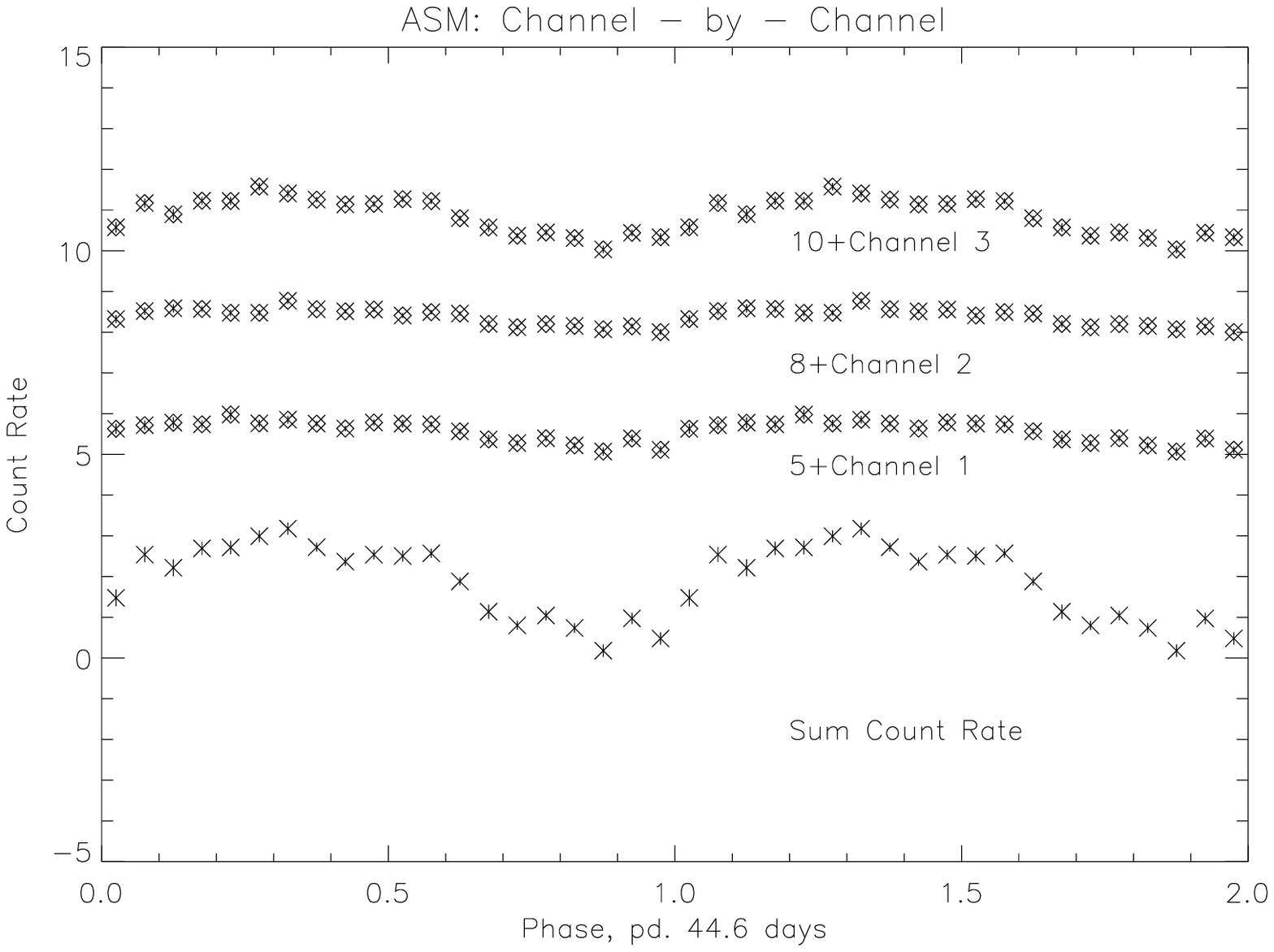,width=8cm,height=8cm}
\caption{ASM total countrate, together with channel - by - channel behaviour, on the same phase scale as figure \ref{fig:fold_hrat}. The low energy channels show smaller amplitude variations and smoother morphology of variation.}                                 
\label{fig:fold_channs}
\end{center}
\end{figure}

When the ASM phase-binned lightcurve is split into its individual
channels (figure \ref{fig:fold_channs}), we see that the amplitude of
variation is highly energy dependent. The variation at energies 1.3 -
3.0 keV, represented by ASM channels 1 and 2, has amplitude $\sim$ 1
$cs^{-1}$ and the variation is smooth, whereas at energies 5.0 - 12.1
keV the variation is about twice this amplitude.

\section{Discussion}

Our analysis of the RXTE and CGRO archival datasets has not only
confirmed the presence of the superorbital period in SMC X-1, it has
clearly demonstrated that this itself is varying on an even longer
timescale (conceivably a ``fourth'' period!).  Here we shall discuss
the superorbital period, along with reasons why it might vary with
time. Four mechanisms have been put forward to explain the
superorbital period, and we shall discuss these for the case of SMC
X-1.

\subsection{Spectral information as a broad discriminant}

Broadly speaking, mechanisms for generating a superorbital period must
do so by a variation in the X-ray intensity of the source, by
variation in uncovered area as seen from the observer, or by a varying
absorption along the line of sight to the observer. However we can
immediately rule out varying absorption. This is because the BATSE
lightcurve varies in the same way as that of the ASM, showing that the
variation occurs across the entire energy range 1.3 - 100 keV. This
rules out absorption by wind or accretion disk. For example, although
the wind density in SMC X-1 can vary by factors of up to $\sim$1000
(Blondin and Woo 1995, Wojdowski et al 2000, see also Vrtilek et al.,
2001), this would have no effect on the BATSE lightcurve. This also
rules out in the case of SMC X-1 the conventional explanation for
superorbital variations, in which varying absorption by a disk warp
causes the periodicity.

In itself, the relation between ASM intensity and spectral hardness
does not act as a good discriminant between mechanisms, as the
lightcurve represents a multiple component source (see Tanaka 1997 for
a review). In the case of neutron star XRB, the soft component
(usually identified with energies $<$ 3 keV) is taken to originate in
the accretion disk, through multi-temperature disk blackbody emission
(Mitsuda et al 1984) combined with a Comptonised component (White,
Stella and Parmar 1988). The wind is also represented at a low level
in the soft component of the emission from SMC X-1
\cite{woj2k}. However it should be noted that a small component of the
soft emission may in some cases arise at the surface of the neutron
star \cite{schul}.

Emission at harder energies is identified with the surface region of
the neutron star itself, through shock or Coulomb heating or Compton
scattering at the boundary layer \cite{fkr}. It has recently been
suggested that the similarity in X-ray spectra between low state
neutron star systems and black hole candidates identifies the dominant
source of the hard component as the advection dominated accretion flow
(ADAF) between the inner disk and compact object \cite{bar}.

Thus the majority of the soft component of the ASM lightcurve can be
identified with the accretion disk or matter ejected from the
companion, whereas the hard component corresponds to emission at or
near the surface of the neutron star itself. In this way, the spectral
dependence of the ASM lightcurve allows us to determine the region in
which variation is taking place. The superorbital variation is
strongly energy dependent: as can be seen from figures
\ref{fig:fold_hrat}, \ref{fig:hrat_counts} \& \ref{fig:fold_channs},
the ASM variation is most pronounced at energies $>$ 5keV, while there
is little or no variation at energies $<$ 3 keV. This suggests we can
identify the observed superorbital variation with changes at or near
the surface of the neutron star.

This interpretation is only valid if ejected material does not
dominate the soft component of the RXTE lightcurve. ASCA spectra of the
X-ray pulsar Cen X-3 \cite{ebi} do indeed contain little or no
variation at energies $<$2 keV in eclipse, whereas the hard component
varies significantly. The authors model the spectrum with three
components (figure 6 of Ebisawa et al 1996): a hard component from the
neutron star itself, significant at energies $>3 keV$, a component due
to electron scattering in circumstellar matter, dominant at energies 1
- 5 keV, and a soft component due to scattering from interstellar
dust, dominant at energies $<2$ keV, by virtue of the location of Cen
X-3 in the galactic plane.

In SMC X-1, however, the circumstellar component cannot be an
important contributor to the ASM lightcurve. In the case of Cen X-3,
the pulse period is strongly energy dependent, with little or no
evidence for the pulsation at energies corresponding to the
circumstellar electron scattering or interstellar dust components
\cite{ebi}, which suggests circumstellar and interstellar scattering
matter are responsible for the emission of the bulk of the soft
X-rays. The position of SMC X-1 out of the galactic plane rules out
any significant interstellar dust component, and its pulse profile is
clear and present right down to 0.1keV \cite{woj}. This rules out
domination of the soft X-ray lightcurve from circumstellar or
interstellar scattering regions. Thus the absence of superorbital
variation in the soft energy bands of the ASM lightcurve is truly
showing that radiation from the disk is not an important participant
in the variation.

\subsection{Mechanisms of Superorbital Variation}

Precession of the neutron star's magnetic axis was invoked to explain
the $\sim$35d periodicity in Her X-1 (by changing the accretion
geometry; see Tr{\"u}mper et al. 1986). However this mechanism can be
ruled out in the case of SMC X-1 for two reasons. Firstly, the
precession torque is expected to be of the same order as the spin-up
torque \cite{lam}. But pointed X-ray observations of SMC X-1 show the
spin period to be monotonically decreasing \cite{woj}, suggesting that
precession should either also monotonically decrease in period, or if
the sign of the torque changes, produce more violent behaviour in the
dynamic power spectrum over time. This is not observed. Secondly, the
precession of the neutron star would result in a superorbital
variation of the pulse profile, whereas pointed observations suggest
that the pulse profile is in fact stable \cite{woj}. Forced precession
has recently been ruled out in the case of Her X-1 for exactly this
reason \cite{scott}.

Another possibility is variation in mass transfer from donor to
neutron star on the superorbital period. Giant pulsators, most notably
RR Lyrae variables, are known to change their pulsation periods (see,
e.g: Jurscik 2001). Thought to be driven by horizontal branch
evolution, these pulsation period changes are typically monotonic and
of order 1 d per 0.1 GYr \cite{weh}. However there are cases in which
the variation is more chaotic and on a faster timescale \cite{jur}, so
we must address this mechanism. The fact that such pulsations have
never been detected in the donor of SMC X-1, given the high
sensitivity of searches for such behaviour, leads us to strongly doubt
the existence of pulsations at all. Furthermore, we expect companion
star pulsations to produce large changes in bulk mass transfer rate
from donor to accretion disk, manifesting themselves as large changes
in the magnitude, or even direction (see, e.g: Nelson et al. 1997) of
spin of the neutron star. As noted by Wojdowski et al (1998), however,
the changes in neutron star spin rate are always in the same sense.

More gentle variation in mass transfer rate can be brought about if
the XRB is part of an hierarchical triple. The third body causes the
inner XRB to undergo nodal precession, leading to mass transfer
variations on a timescale determined by the orbital periods of the
binary and hierarchical third body \cite{chg}. This mechanism could
give rise to both the superorbital period and the longer period of its
variation, with the orbit of the third body the superorbital period
and its variation period that of the nodal precession. Such a
mechanism can be ruled out for SMC X-1, however, by consideration of
the orbit of the centre of mass of the inner XRB about the system
centre of mass, and the variation in pulse arrival time this would
produce. Measured discrepancies between pulse arrival delays and those
predicted from a two-body system amount to less than $\sim$ 10ms in
RXTE PCA data \cite{woj}. To produce such a small variation in pulse
arrival time, the third body would have to orbit the inner XRB with
the normal vector of the orbit pointing $<$ 2\degr from the line of
sight, for any stellar third body. Thus a third body scenario requires
an extraordinary degree of fine-tuning to reproduce the observed
results, especially when we consider that SMC X-1 itself is an
eclipsing binary and thus would have high orbital inclination with
respect to the orbit of the third body.

This leaves us with precession of a warped accretion disk as the
mechanism responsible for the superorbital period. Such systems are
usually thought to have as their primary observational effect a change
in absorption, which has been ruled out for this system. What is not
ruled out is a variation in uncovered emitting area, as suggested for
Her X-1 (Gerend \& Boynton 1976). Also not ruled out is variable
deposition of energy at the boundary layer, brought on by the
quasi-steady nature of the precession. This is not in conflict with
the spin-up of the neutron star, as there are deviations from a purely
monotonic increase \cite{woj}. Indeed if one looks at the residuals in
the quadratic fit to the pulse period \cite{woj}, the residuals show
variation of similar shape and timescale to the variations of the
superorbital X-ray period! It is currently uncertain if this is a real
effect or a statistically insignificant artifact. If statistically
significant, this would provide direct evidence linking the variation
in third period to variation in accretion flow onto the neutron star.

There is a third way in which a warped, precessing accretion disk
might manifest itself in the X-ray lightcurves examined here. Studies
of CVs have shown that local values of mass transfer in accretion
disks can dramatically exceed the mass transfer rate from donor to
acceptor, as shown in CV's with prominent bright spots such as LX Ser
\cite{rvp}. In the case of SMC X-1, the quasi-steady decrease in spin
period noted by Wojdowski et al (1998) suggests that mass transfer
from the donor star is also quasi-steady, either in a Roche Lobe
stream or collimated wind, allowing such a bright spot to exist. SMC
X-1 is expected to have a small accretion disk given its HMXB nature,
which, if warped, might allow the mass transfer stream to reach close
enough to the neutron star that the resulting bright spot becomes an
important component in the X-ray emission of the system, as suggested
by Warner (2002, private communication). As the warp precesses, the
intersection point moves radially through the disk, thus varying the
brightness and spectral hardness of emission - as seen in figure
\ref{fig:fold_hrat}. Furthermore, as the mass transfer stream would be
varying its position on the precession period, the intersection point
would drive the warp further. Such a mechanism is deserving of further
theoretical investigation.

The warped, precessing accretion disk has one clear advantage over the
other mechanisms considered here: it can support variations in
superorbital period. To show this, we summarise first the properties
of warped accretion disks.

\subsection{Stability of Precessing, Warped Accretion Disks}

The hypothesis that warped, precessing accretion disks might be
responsible for superorbital periods in XRB gained acceptance through
the canonical precessing disk systems Her X-1, SS 433 and LMC X-4
\cite{jp}. The measured $\sim$160d precession of the relativistic jets
in SS433 have also been identified with precession of the accretion
disk \cite{mar}.
Tidal forces on the disk by the companion can lead to a constant rate
of precession if the disk can become tilted out of the plane (see eg:
Wijers and Pringle 1999), but two issues then arise: (i) how to cause
a fluid disk to precess with uniform frequency, and (ii) how to
maintain tilting out of the plane.

Modelling of disk behaviour by Wijers \& Pringle \shortcite{wp} and
subsequent more rigorous treatment by Ogilvie \& Dubus (2001:
hereafter OD01) has shown that an accretion disk is unstable to
radiation-driven warping when the luminosity of the central source
exceeds a critical value. This instability provides a mechanism
enabling the disk to tilt and to remain tilted and so to precess with
a period that is comparable to the recorded values for superorbital
periods in X-ray binaries. Furthermore the models also provide for the
possibility of both retrograde and prograde motion.  Thus all the
superorbital periods seen in X-ray binaries might now be accountable
by the same model.

The most pertinent of the OD01 results for this discussion is the
classification of behaviour into modes of warping (see figure 7 of
OD01). The stability of the accretion disk to warping is predicted as
a function of the mass ratio $q$ and the binary separation $r_b$, for
two regimes of mass input. The first region boundary corresponds to
matter input at the Lindblad resonance radius $r_o$, for which every
source above this line for which matter is input in this way can
sustain mode 1 and higher warps. If matter is input at the
circularisation radius $r_c$, this border does not apply, and
persistent warping is not possible. Between the upper border of this
region and the dotted line, mode 0 warping persists. Finally in the
uppermost region, mode 1 and higher warping modes become possible.

The location of the borders between stability regions in $r_b$ - $q$
space depends strongly on the global disk viscosity parameter,
$\alpha$ and the neutron star accretion efficiency $\eta$. Uncertainty
in these values translates into uncertainty as to the location of the
border between regions and thus the stability classification of a
system. However, the adopted values of $\alpha$ = 0.3 and $\eta$ = 0.1
are usually assumed to be appropriate for neutron star XRB (Frank,
King \& Raine 1995, OD01 and references therein).

As a HMXB, the accretion disk in SMC X-1 is unlikely to reach out to
$r_o$. Moreover, that the spin period has been decreasing in a steady
way compared to most other HMXB's suggests that mass transfer from
donor to disk is also comparatively steady, suggesting mass transfer
is collimated in some way, even if not by Roche overflow. These
considerations lead us to suggest that the true site of mass input is
closer to $r_c$ than to $r_{o}$, so we use the stability scheme
appropriate for mass input at $r_c$. SMC X-1 is then placed near the
border between the region supporting stable mode 0 warping and that
supporting mode 1 and higher modes. We would thus expect the disk to
form a precessing warp, giving rise to a long periodicity. However,
this may not be a stable monotonic warp; instead, there may be an
interaction between modes by virtue of the unique location of SMC X-1
near the border between the two regions. As two or more modes compete,
then, a longer term variation may be set up in which the sum of modes
itself changes slowly with time. This may manifest itself as the
observed variation in the superorbital period of SMC X-1, which must
represent an important constraint on future development of such
models.

\section{Conclusion}

We have used the dynamic power spectrum to show that the superorbital
period of SMC X-1 varies in a coherent, apparently almost sinusoidal,
way. In conjunction with the spectral behaviour of this variation, we
have seen that a precessing, warped accretion disk must be responsible
for the superorbital variation. However the influence of the warp may
be felt not only through varying occultation, as is supposed for Her
X-1, but through varying accretion at the neutron star boundary layer,
permitted because the disk precession is not steady but
quasi-steady. Finally we saw that the variation of the superorbital
period is fully consistent with current stability theory of warps in
accretion disks. An interaction of warp modes can give rise to the
varying superorbital period, allowing its variation to change
direction several times in a few years.

This remarkable result was only possible due to the superb, long
time-base of the CGRO and RXTE databases.  It heralds a new
understanding of the structure and evolution of accretion disks which
may have implications for disks on all scales.  A search for similar
behaviour in other X-ray binaries in these datasets is under way and
will be the subject of future papers.

\bigskip

{\bf Acknowledgments}

SGTL and WIC were in receipt of PPARC research studentships. WIC
thanks Professor Brian Warner for interesting discussion on the
observational consequences of tilted and warped disks, and Drs. Phil
Uttley and Guillaume Dubus for informative comments. This work was
only made possible through the efforts of the ASM/RXTE team at MIT and
GSFC, and the BATSE/CGRO team at MSFC.

\label{lastpage}

\end{document}